# File System in Data-Centric Computing

Viacheslav Dubeyko, *Western Digital Research*

## 1. What Is Data-Centric Computing?

The CPU-centric model means that data lives in persistent storage and it needs to copy data from persistent storage into DRAM with the goal to process data by means of central computing engine. The CPU-centric model is based on von Neumann architecture that revealed a lot of drawbacks (such as the bandwidth wall).

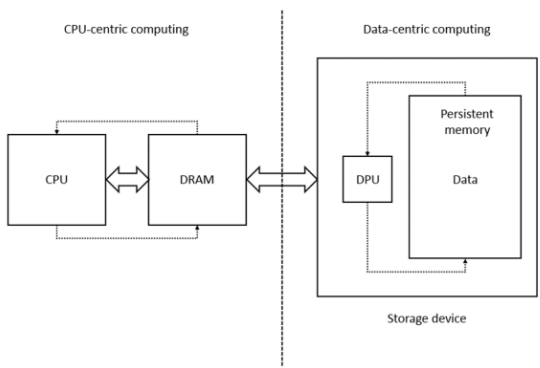

**Fig. 1 CPU-centric vs. Data-centric computing**

Oppositely, the data-centric computing or near-data processing (NDP) model implies that data lives in the persistent storage with processing engines surrounding the data and transforming data in-place. The Big Data problem requires completely new approaches that will be able to improve the performance and to remove the drawbacks of mature and robust concepts of data storage and the file system. The data-centric computing is able to: (1) exclude the extensive exchange by user data between the host and storage device; (2) overcome the problem of DRAM's expensiveness and limited capacity for the case of Big Data processing; (3) achieve the native concurrency of data processing.

## 2. Related Research Works

There are a lot of research efforts in the academia that are dedicated to the Near-Data Processing approach. It's worth to mention such implementations like Active Flash [1], Intelligent SSD [2], Smart SSD [3], Collaborative in-SSD processing [4], XSD [5], Active Disk/iSSD [6], Minerva [7], BlueDBM [8], PRINS [9]. All these research works claim significant improvement of data processing performance. Generally speaking, Near-Data Processing in storage sounds like very promising paradigm.

## 3. "Smart" Storage Device like a Trend

The Intel Movidius Neural Compute Stick [10] is one of the example of the moving computation on the edge. It is possible to expect a lot of different "smart" storage devices with different features in the near future. The moving computation on the edge or near to data is the new trend that can break the bandwidth wall and to unleash the power of next generation NVM/SCM memory.

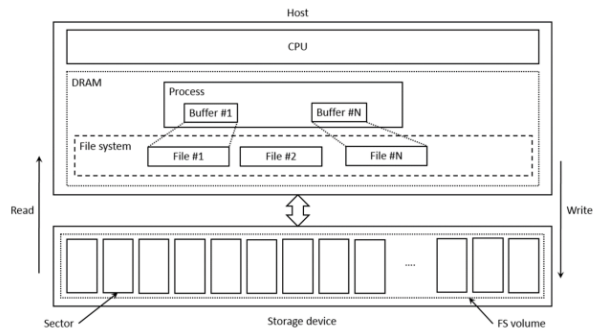

**Fig. 2 File system in CPU-centric computing**

## 4. File System Concept

File system is the important OS subsystem that plays the role of mediator between the user-space application and storage device. The key goal of the file system is to represent the file abstraction and to build the files' namespace. Namely this concept provides the way of access and modifies the user data on the host side. Nowadays, the file system provides the fundamental interface: (1) open a file in the process/thread; (2) read/write some portion of file's content; (3) sync/close the opened file.

## 5. File System in Data-Centric Computing

However, the file system can play efficiently for the case of data-centric computing. In the current paradigm the file system needs to copy the metadata and user data in the DRAM of the host with the goal to access and to modify the user data on the host side. The DAX approach doesn't change the concept but to build the way to bypass the page cache via the direct access to file's content in persistent memory. Generally speaking, for the case of data-centric computing, the file system needs to solve the opposite task not to copy data into page cache but to deliver the processing activity near data on the storage device side.

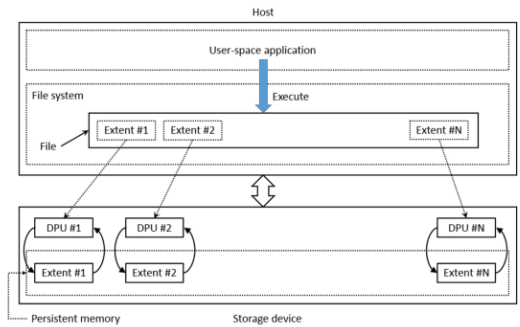

**Fig. 3 File system in data-centric computing**

And file system looks like the proper place in the whole stack to take the responsibility to manage data processing on the data storage side.

## 6. Data Abstraction in Data-Centric Paradigm

Usually, file system treats the user data like a byte stream (no knowledge about user data format). However, in the case of data-centric computing, it is possible to treat any file like a container that includes a sequence of items or extents. This concept provides the opportunity to implement the generalized algorithms of data processing on the storage device side (likewise C++ standard library). The responsibility of user-space application will be to store a user data as sequence of items (extents) and to provide specialized implementation of method that can be used by generalized algorithms on the storage device side.

## 7. Interface of User-Space Application with File System in Data-Centric Computing

Existing interface of interaction a user-space application with file system can be extended by: (1) OPEN – to open file with the support of data-centric computing; (2) FREEZE – to freeze some portion(s) or the whole file for data processing on the storage device side; (3) GET – to find file's items that correspond to the requested condition; (4) READ – to extract the found items to the host side; (5) WRITE – to store persistently the result of the GET operation; (6) SET – to modify on the storage device side the file's items that correspond to the requested condition; (7) EXECUTE – to apply an operation on the storage device side for one or group of files.

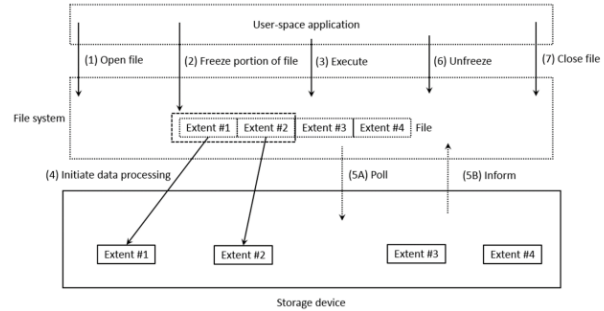

**Fig. 4 Data processing workflow**

User space application is able to initiate data processing on the storage device side in several possible modes: (1) SYNCHRONOUS mode – host's thread tracks the ending of operation on storage device side via polling or interrupt technique; (2) ASYNCHRONOUS mode – host delegates the processing activity on the storage device side by placing requests in a storage device's queue till the queue's contention; (3) DELAYED EXECUTION mode – host stores persistently data processing requests on storage device side and the storage device processes the requests internally; (4) TRIGGER mode – storage device has registered methods that react on some events (add data, modify data, delete data and so on).

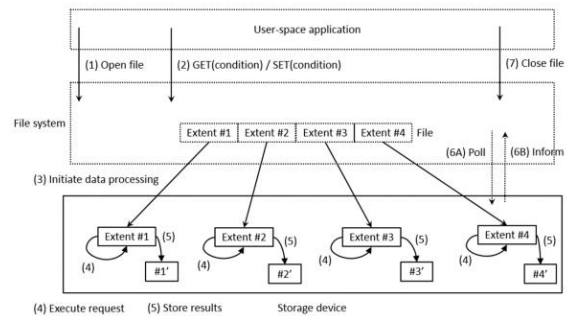

**Fig. 5 GET/SET operation**

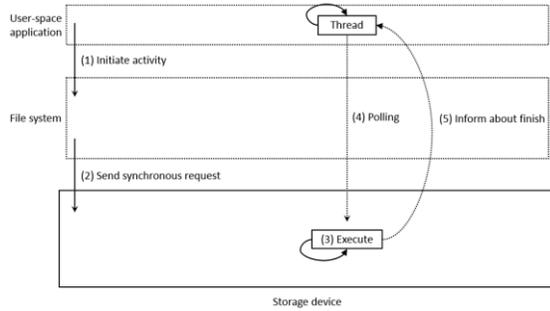

**Fig. 6 Synchronous mode**

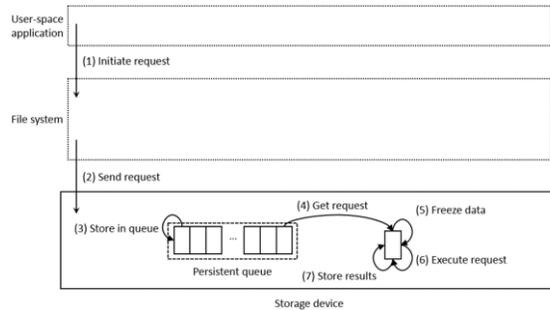

**Fig. 7 Delayed execution mode**

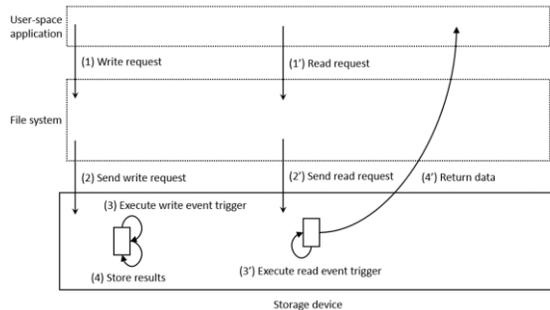

**Fig. 8 Trigger mode**

The distribution of data processing activity between the host and the storage device can be done by user space application itself or by compiler during the compilation process.

## 8. References

[1] S. Boboila, Y. Kim, S. S. Vazhkudai, P. Desnoyers and G. M. Shipman, "Active Flash: Out-of-core data analytics on flash storage," 012 IEEE 28th Symposium on Mass Storage Systems and Technologies (MSST), San Diego, CA, 2012, pp. 1-12. (https://ieeexplore.ieee.org/document/6232366/)

[2] Duck-Ho Bae, Jin-Hyung Kim, Sang-Wook Kim, Hyunok Oh, and Chanik Park. 2013. "Intelligent SSD: a turbo for big data mining," In Proceedings of the 22nd ACM international conference on Conference on information & knowledge management (CIKM '13). ACM, New York, NY, USA, 1573-1576. (http://elib.mi.sanu.ac.rs/files/journals/csis/38/csisn38p375-394.pdf)

[3] Y. Kang, Y. s. Kee, E. L. Miller and C. Park, "Enabling cost-effective data processing with smart SSD," 2013 IEEE 29th Symposium on Mass Storage Systems and Technologies (MSST), Long Beach, CA, 2013, pp. 1-12. (https://ieeexplore.ieee.org/document/6558444/)

[4] Yong-Yeon Jo, SungWoo Cho, Sang-Wook Kim, and Hyunok Oh. 2016. Collaborative processing of data-intensive algorithms with CPU, intelligent SSD, and GPU. In Proceedings of the 31st Annual ACM Symposium on Applied Computing (SAC '16). ACM, New York, NY, USA, 1865-1870. (https://dl.acm.org/citation.cfm?id=2851741)

[5] Cho, Benjamin Y., Won Seob Jeong, Doohwan Oh and Won Woo Ro. "XSD : Accelerating MapReduce by Harnessing the GPU inside an SSD." (2013). (https://pdfs.semanticscholar.org/adcd/426b1235d9c68ab0432867bcc02a661b8be1.pdf)

[6] Sangyeun Cho, Chanik Park, Hyunok Oh, Sungchan Kim, Youngmin Yi, and Gregory R. Ganger. 2013. Active disk meets flash: a case for intelligent SSDs. In Proceedings of the 27th international ACM conference on International conference on supercomputing (ICS '13). ACM, New York, NY, USA, 91-102. (http://www.pdl.cmu.edu/PDL-FTP/Storage/CMU-PDL-11-115.pdf)

[7] Arup De, Maya Gokhale, Rajesh Gupta, and Steven Swanson. 2013. Minerva: Accelerating Data Analysis in Next-Generation SSDs. In Proceedings of the 2013 IEEE 21st Annual International Symposium on Field-Programmable Custom Computing Machines (FCCM '13). IEEE Computer Society, Washington, DC, USA, 9-16. (https://cseweb.ucsd.edu/~swanson/papers/FCCM2013Minerva.pdf)

[8] Sang-Woo Jun, Ming Liu, Sungjin Lee, Jamey Hicks, John Ankcorn, Myron King, Shuotao Xu, and Arvind. 2015. BlueDBM: an appliance for big data analytics. In Proceedings of the 42nd Annual International Symposium on Computer Architecture (ISCA '15). ACM, New York, NY, USA, 1-13. (http://livinglab.mit.edu/wp-content/uploads/2016/01/ISCA15_Sang-Woo_Jun.pdf)

[9] Leonid Yavits, Roman Kaplan, Ran Ginosar, "PRINS: Resistive CAM Processing in Storage," 2018. (https://arxiv.org/abs/1805.09612)

[10] Intel Movidius Neural Compute Stick. (https://developer.movidius.com/)